\documentclass[twocolumn,rmp,amsmath,amssymb,showpacs]{revtex4}
\usepackage{graphicx}
\usepackage{dcolumn}
\usepackage{bm}

\begin{document}

 \title{Ageing effects around the glass and melting transitions in poly(dimethylsiloxane) visualized by resistance measurements}

 \author{H. B. Brom}

 \author{I. G. Romijn}

 \author{J. G. Magis}

 \affiliation{Kamerlingh Onnes Laboratory, Leiden University, P.O. Box 9504, 2300 RA Leiden, The Netherlands}

 \author{M. van der Vleuten}\altaffiliation{also project student at Oc\'{e} Technologies,  Venlo, The Netherlands}

 \author{M. A. J. Michels}\altaffiliation{also at the Dutch Polymer Institute Eindhoven, The Netherlands}

 \affiliation{Group Polymer Physics, Technische Universiteit Eindhoven, P.O. Box 513, 5600 MB Eindhoven, The Netherlands}

 \date{July 27 2004, Appl. Phys. Lett. {\bf 85}, 570 (2004)}

\begin{abstract}
The process of ageing in rubbers requires monitoring over long
periods (days to years). To do so in non-conducting rubbers, small
amounts of carbon-black particles were dispersed in a fractal
network through the rubber matrix, to make the rubber conducting
without modifying its properties. Continuous monitoring of the
resistance reveals the structural changes around the glass and
melting transitions and especially details about the hysteresis
and ageing processes. We illustrate the method for the
semicrystalline polymer poly(dimethylsiloxane) (PDMS).
\end{abstract}


\maketitle

The properties of amorphous polymers (AP) often change in time
below the rubber-glass transition temperature $T_{\rm
g}$\cite{Strobl97}. For $T<T_{\rm g}$ the material can be regarded
as a solidified supercooled liquid of which the volume, enthalpy
and entropy are larger than in the equilibrium state
\cite{Struik87,Hodge95}. The gradual approach to equilibrium
affects various properties of the system (not only mechanical,
like elasticity, creep- and stress-relaxation rates, but also
dielectric and optical \cite{Cheun04}), and is called physical
ageing to distinguish it from chemical influences. For many
plastics or rubbers the ageing range includes the temperatures for
practical application. The same processes are playing a role in
semi-crystalline polymers (SCP), like poly(dimethyl siloxane)
(PDMS), but in addition the presence of very thin crystal lamellae
makes them inhomogeneous, giving a strong resemblance to
composites \cite{Struik87}. But unlike the amorphous rubbers,
semi-crystalline polymers turn out to age at temperatures above
their $T_{\rm g}$. Usually these changes are monitored via
mechanical probes (elasticity measurements), or in case of polar
constituents also spectroscopically. For non-conducting non-polar
crystalline and amorphous polymers addition of Carbon-Black (CB)
particles has been used to make these material conducting, and the
conductivity and noise data were shown to have marked relaxation
effects \cite{Klason75}. Here we revisit this approach and
demonstrate that properly dispersed small amounts of conducting CB
particles form a {\it fractal} network through the polymer, of
which the conductivity properties can be well understood
\cite{Adriaanse97}. The CB particles do not influence the rubber
and glass transition in an essential way and hence the time
evolution of the conductivity is representative for changes in the
polymer itself. We illustrate the method for PDMS. The data turn
out to be not only a very sensitive detector for subtle changes
due to the rubber and glass transitions as found previously
\cite{Klason75}, but also demonstrate that continuous monitoring
of the resistance gives insight in the long time constants of the
aging process.

PDMS has a $T_{\rm g}$ of the amorphous phase around 150~K and a
melting transition temperature $T_{\rm m}$ of the crystalline
phase at 222~K \cite{Strobl97,Dollase02}. This difference of 70~K
makes the material very convenient to study the influence of both
transitions on the polymer structure below room temperature. The
PDMS chains have a typical length of hundred repeat units, i.e.
about 100~nm, and are crosslinked at the end points only. As
conducting filler material CB is attractive as it is known to form
fractal distributions \cite{BundeHavlin91} with percolation
thresholds of the order of a few tenths of a percent or lower in
volume fraction, when dispersed under the proper circumstances
\cite{Adriaanse97}. The CB (Printex XE-2 from Degussa, Probst) in
the present study is quite similar to the Ketjenblacks used
previously \cite{Adriaanse97}. The primary particles are hollow
half spheres with a typical diameter of 30~nm, which in the gas
phase have formed aggregates of 100~nm. The sample quality was
checked by confocal scanning and optical microscopy.

\begin{figure}[h,t]
 \begin{center}
 \includegraphics[width=5cm]{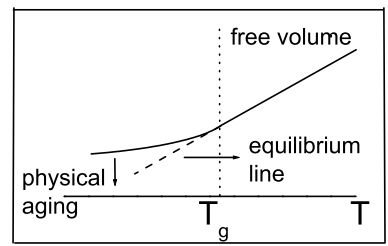}
 \end{center}
 \caption{Relation between $T_{\rm g}$ and $V_{\rm f}$ for an AP. During ageing,
 which typically occurs below $T_{\rm g}$, $V_{\rm f}$ is reduced.\label{fig.1}}
 \end{figure}

As shown by Struik \cite{Struik87}, in most of its aspects ageing
in AP and SCP can be explained in a straightforward way from
free-volume concepts ($V_{\rm f}$), see Fig. 1. The segmental
mobility of the polymer chains in a dense system is primarily
determined by the free volume, and above a critical degree of
packing steeply falls to zero.

Below the glass temperature, the AP and SCP will behave similarly,
but as mentioned already, unlike AP the SCP turn out to age above
their $T_{\rm g}$, for the following reason. Let us consider a SCP
above $T_{\rm g}$. Since the amorphous polymer chains are
connected to the crystallites, the segmental mobility near the
particle surface will be reduced. Only far away from the
crystallites, the mobility and other properties of the rubber
matrix will be equal to those of the amorphous rubber. The glass
transition of a SCP will be broadened on the high temperature
side. By ageing, which will now occur in a wide range of
temperatures above $T_{\rm g}$, the amorphous regions will become
smaller (a clear shift in $T_{\rm g}$ will only be seen if the
amorphous phase is completely disturbed). It means that there are
three regions to discriminate: above $T_{\rm m}$ the polymer is
completely amorphous and below $T_{\rm g}$ it has become quite
rigid. In between these two temperatures the rubber has the
possibility of (re)crystallization and especially the
crystallization will crucially depend on the thermal history.

The electrical conduction in systems with a low concentration of
conducting particles dispersed in a non-conducting matrix, is well
documented and occurs via Mott variable-range hopping. The
$T$-dependent conductivity can be written as $\sigma = \sigma_0
\exp [-(T_0/T)^{\gamma}]$, where the value of $\gamma$ depends on
details of the filler distribution, the filler density of states,
and the importance of Coulomb interactions
\cite{BundeHavlin91,Adriaanse97}.

We performed two kinds of measurements: temperature scans
(dependence of the resistance $R$ on $T$ and cooling/warming rate
$dT/dt$) and time scans (time evolution of $R$ at constant $T$).
In the latter we compared results of a 'down-quench' $T_{\rm room}
\downarrow T$ with those of a 'down-up-quench' $T_{\rm room}
\downarrow 10~{\rm K} \uparrow T$, where we discriminate between
three regimes for the final temperature: $T < T_{\rm g}$, $T_{\rm
g} < T < T_{\rm m}$ and $T_{\rm m} <T$. The (4-points) data were
also taken with pulsed fields with different field strengths to
check for ohmic behavior while avoiding possible heating. Data
taken at room temperature as function of filler volume fractions
are well represented by $\sigma(p) = \sigma_0 (p - p_c)^t$ with $t
= 2.3 \pm 0.2$ and $p_c = 0.0033 \pm 0.0004$. In
Figs.~{\ref{TvsP},\ref{RvsT}a we show the resistivity as function
of temperature $T$.

\begin{figure}[h,b]
\begin{center}
\includegraphics[width=7.5cm]{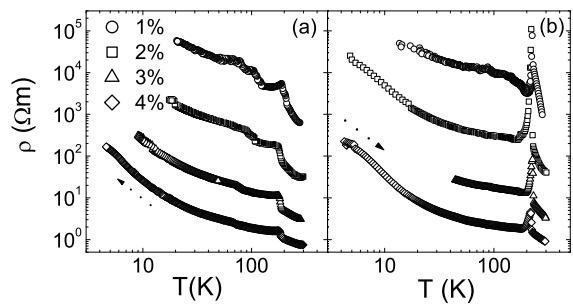}
\end{center}
\caption{The $T$ dependence of the resistivity $\rho$ for various
filler concentrations $p$. The value of $p$ is seen to affect only
the absolute resistance values. The data while cooling down (a) or
heating up (b) are clearly different, see also the next figure.
\label{TvsP}}
\end{figure}

\begin{figure}[h,t]
\begin{center}
\includegraphics[width=7.5cm]{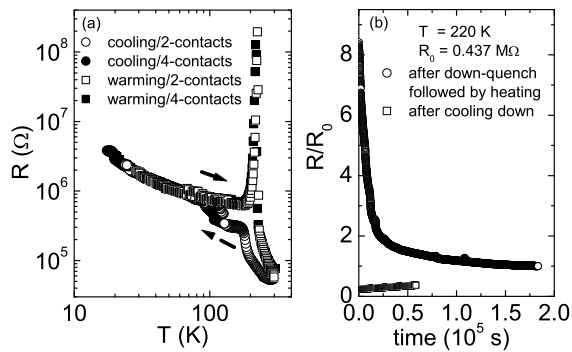}
\end{center}
\caption{(a) Hysteresis in $R$ vs. $T$ for $p=0.15$. Cooling-down
($\downarrow$) and warming-up ($\uparrow$) rates were a few Kelvin
per minute. Data taken with 2 and 4-contacts (separate current and
voltage contacts) were identical (differences at 200~K in the
$\downarrow$ data are due to differences in cooling rate). Around
220~K the maximal $\uparrow$ and $\downarrow$ resistances differ
by more than 2 orders of magnitude. (b) Growth and decrease of the
normalized resistance versus time close to $T_{\rm m}$ during
cooling down from 300~K, resp. warming up from 10~K. The
resistances are both normalized to 0.437 M$\Omega$, the value
reached after stabilizing the temperature for 2 days at 220~K
after warming up from 10~K. \label{RvsT}}
\end{figure}

{\it $R$ as function of $T$.} When we compare cooling-down and
warming-up curves, the data only merge below 80~K and above 320~K,
and show strong hysteresis in between. Apparently the region
between 320~K and 80~K is affected by the  melting and
rubber-glass transitions, which from Differential Scanning
Calorimetry data for all studied rubbers were found to be around
220~K and 160~K respectively, in agreement with the literature
\cite{Strobl97}. Dynamical Mechanical Analysis (DMA) of the same
systems revealed the rubber-glass transition as a peak in $\tan
\delta$ at 154~K and the melting temperature around 220~K by a
loss in viscosity. Especially because the resistance data show the
same pattern independent of CB filler fractions, we conclude that
at these small concentrations the CB particles do not change the
polymer properties in an essential way and the aging effects in
$\sigma$, which are in general due to subtle changes in the
particle network, are caused by the polymer matrix. As in previous
CB samples \cite{Adriaanse97}, sufficiently far below the glass
transition the conductivity is well described by $\sigma =
\sigma_0 \exp [-(T_0/T)^{\gamma}]$. From fits to $\sigma$ between
20 and 80~K for the 1.5 \% sample we find $\gamma = 0.68$ and $T_0
= 50 \pm 7$~K. For the interpretation we refer to
Ref.~\onlinecite{Adriaanse97}, where these values (which can be
seen as a measure for the quality of the dispersion) were argued
to be characteristic for percolation on a fractal carbon network.

{\it $R$ as function of time $t$}. For $T_{\rm m} < T < 300$~K,
the resistance after a down-quench remained constant (measured up
to $10^5$~s, not shown). The same is true for $T < T_{\rm g}$.
Strong time dependences where seen for $T_{\rm g} <T<T_{\rm m}$,
for which a typical result is shown in Fig.~\ref{RvsT}b.
Down-up-quench results with a final temperature below $T_{\rm g}$
do not show noticeable changes, which is also the case for the
down-quench data. For $T > T_{\rm m}$ the similarity between the
two kinds of measurements is absent and long-time tails become
visible after heating, see Fig.~\ref{RvsT2}. In the region $T_{\rm
g} < T < T_{\rm m}$ the time constants are again quite long, see
Fig.~\ref{RvsT}b.

\begin{figure}[h,b]
\begin{center}
\includegraphics[width=7.5cm]{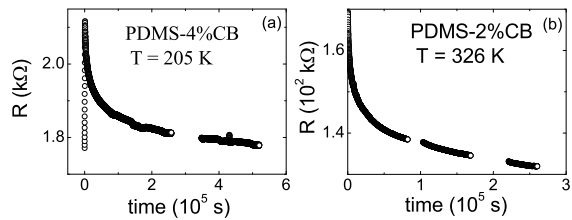}
\end{center}
\caption{(a) Increase and decrease of $R$ versus $t$ after a
down-quench from 300~K to 205~K. (b) Decrease of $R$ versus $t$
after first cooling down from room temperature to 15~K and
thereafter warming up to 326~K. \label{RvsT2}}
\end{figure}

How can we explain the overall features of the resistivity versus
temperature and what is their connection with the melting and glass
transition?

The increase in $R$ with cooling is a general feature for hopping
transport, where the available phonon energy plays an important
role. The melting point remains almost unnoticed upon cooling the
sample (typical cooling rate of 1-3 K per minute). This is likely
because formation of crystallites is delayed and hence does not
affect the distances in the percolation path. Things changes
around 190~K, where crystallization seems to nucleate and starts
to change the CB distances and the segmental mobility. By this
process some amorphous regions might even pass through the glass
transition (see also below), by which the structural
reorganization process will be slowed down appreciably. The $R$
versus $T$ curves below 80~K can be explained in the same way as
in Ref.~\onlinecite{Adriaanse97}.

During heating from about 10~K the warming-up resistance curves
remains rather smooth above 80~K, deviating strongly from the
cooling down data. Some reorganization seems to occur, a process
that becomes more important between $T_{\rm g}$ and $T_{\rm m}$.
Because of the increase in chain mobility further crystallization
is possible which leads to more free volume, etc. We believe that
this process is at the heart of the resistance anomaly around the
melting temperature of 220~K. Above $T_{\rm m}$ the crystals melt
and the whole polymer system becomes amorphous. Note that, in
contrast, in the DMA data  mainly the rubber-glass transition
shows up during warming up. This phenomenon might find its
explanation in the specific character of the conductance data,
which are very sensitive to changes on small (atomic) length
scales, while DMA and other methods that are applied from outside
probe mean values where longer distances are involved.

What can we learn more by looking at the time dependence of $R$?
The fact that after a down-quench to $T$ above $T_{\rm m}$ the
resistance remains stable, is no surprise as no isothermal volume
relaxation or crystallization is expected: the system is still in
thermodynamic equilibrium. The effect of the down-quench to 205~K
is remarkable (Fig.~\ref{RvsT2}a). First the resistance grows and
thereafter it decreases with a much longer time constant. It also
appears that the closer we go to the melting temperature the
longer the initial time constant becomes. Most likely the increase
in resistance is due to the formation of crystalline regions. This
process will redistribute the CB particles and create some longer
distances and hence so called "hard hops" in the percolation path
through the sample, which will increase the resistance. In the
neighborhood of the crystalline material the chain mobility is
reduced - some amorphous regions are passing through their glass
transition and therefore we are now in the broadened $T_{\rm g}$
regime, with slow isothermal volume relaxation. The removal of
free volume will bring the CB particles closer together and lead
to a decrease in $R$.

In summary, dispersing small amounts of CB particles in a {\it
fractal} network in non-conducting rubbers opens a convenient way
for continuously monitoring the structural changes on small length
scales due to ageing. Even above the melting temperature we
observe long time constants in the relaxation and therefore have
to conclude that the sample is still not in thermal equilibrium.
Only by waiting for several days, the hysteresis loop seems to
close again.

\indent We acknowledge Oc\'{e} Technologies for providing the
samples and hospitality (MvV), and Sander Smit and Iulian Hulea
for participation in the measurements. The work of I.G.R. forms
part of the research programme of FOM-NWO.


\begin{thebibliography}{1}
\bibitem{note} At the same time, although $\sigma$ is due to the
CB-particles, its changes are driven by the polymermatrix.
\end{thebibliography}
\end{document}